\documentclass[aps,twocolumn,prl,superscriptaddress,longbibliography,reprint]{revtex4-2}
\usepackage{amsfonts}
\usepackage{mathrsfs}
\usepackage{amsmath}
\usepackage{color}
\usepackage{graphicx}
\usepackage{bm}
\usepackage{amssymb}
\usepackage{xspace}
\usepackage{epstopdf}
\usepackage{longtable}
\usepackage{multirow}
\usepackage{booktabs}
\usepackage{textcomp}
\usepackage[title]{appendix}
\usepackage{float}
\usepackage{dsfont}
\usepackage[colorlinks=true, letterpaper=true, pdfstartview=FitV, linkcolor=blue, citecolor=blue, urlcolor=blue]{hyperref}
\usepackage[]{changes}
\usepackage{overpic}

\providecommand{\U}[1]{\protect\rule{.1in}{.1in}}

\begin{document}

\title{Intrinsic Nonlinear Spin Hall Effect and Manipulation of Perpendicular Magnetization}

\author{Hui Wang}
\thanks{These authors contributed equally to this work.}
\affiliation{Division of Physics and Applied Physics, School of Physical and Mathematical Sciences, Nanyang Technological University, Singapore 637371, Singapore}

\author{Huiying Liu}
\thanks{These authors contributed equally to this work.}
\affiliation{School of Physics, Beihang University, Beijing 100191, China}

\author{Xukun Feng}
\thanks{These authors contributed equally to this work.}
\affiliation{Research Laboratory for Quantum Materials, Singapore University of Technology and Design, Singapore 487372, Singapore}

\author{Jin Cao}
\affiliation{Institute of Applied Physics and Materials Engineering, Faculty of Science and Technology, University of Macau, Macau SAR, China}

\author{Weikang Wu}
\affiliation{Key Laboratory for Liquid-Solid Structural Evolution and Processing of Materials, Ministry of Education, Shandong University, Jinan 250061, China}

\author{Shen Lai}
\affiliation{Institute of Applied Physics and Materials Engineering, Faculty of Science and Technology, University of Macau, Macau SAR, China}

\author{Weibo Gao}
\email{wbgao@ntu.edu.sg}
\affiliation{Division of Physics and Applied Physics, School of Physical and Mathematical Sciences, Nanyang Technological University, Singapore 637371, Singapore}

\author{Cong Xiao}
\email{xiaoziche@gmail.com}
\affiliation{Institute of Applied Physics and Materials Engineering, Faculty of Science and Technology, University of Macau, Macau SAR, China}

\author{Shengyuan A. Yang}
\affiliation{Institute of Applied Physics and Materials Engineering, Faculty of Science and Technology, University of Macau, Macau SAR, China}

\begin{abstract}
We propose an intrinsic nonlinear spin Hall effect,
which enables the generation of collinearly-polarized spin current in a large class of nonmagnetic materials with the corresponding linear response being symmetry-forbidden. This opens a new avenue for field-free switching of perpendicular magnetization, which is required for the next-generation information storage technology. We develop the microscopic theory of this effect, and clarify its quantum origin in band geometric quantities which can be enhanced by topological nodal features. Combined with first-principles calculations, we predict pronounced effects at room temperature in topological metals $\mathrm{PbTaSe_{2}}$ and PdGa. Our work establishes a fundamental nonlinear response in spin transport, and
opens the door to exploring spintronic applications based on nonlinear spin Hall effect.
%
\end{abstract}
\maketitle

Nonlinear transport effects have attracted tremendous attention in recent research~\cite{gao2014,fu2015,facio2018,du2018,ma2019,kang2019,du_disorderinduced_2019,Sodemann2019,Shao2020,Isobe2020,kumar2021room,He2021quantum,Tiwari2021,Zhang2021,lai2021,lu2021,He2022graphene,Liu2022PRL,sinha2022,Liao2023,Gao2023QM,Wang2023QM,Huang2023,Song2023ASK,Ortix2024,Lee2024,Das2024,Han2024room,Xu2024electronic}. They manifest novel band geometric properties, provide new tools to characterize materials, and are promising for applications. Of particular interest is a class of intrinsic effects, which are  determined solely by the band structure and represent properties intrinsic to the material. Several intrinsic nonlinear charge transport effects have been proposed and probed in experiment \cite{gao2014,Yan2020PRR,wang2021,liu2021,Gao2023QM,Wang2023QM,ortix2021,culcer2021,Zhou2022,Cao2023,Wang2024IPHE,Wang2024Exp,Han2024room,Xu2024electronic,Xiang2023third,Huang2023NPHE,Chen2024PRR}. Recent efforts have been extending the study towards nonlinear spin responses~\cite{Nagaosa2017,Hayami2022,zhang2023symmetry,Wang2022,xiao2022,xiao2023,Kodama2024,Baek2024nonlinear}. Notably, the spin Hall effect, a prominent effect extensively studied in the past two decades \cite{Jungwirth2012SHE,Hoffmann2013SHE,sinova2015}, also has a nonlinear counterpart which, from symmetry perspective, should have an intrinsic contribution. Nevertheless, a theory of this intrinsic nonlinear spin Hall effect (NSHE) has not been formulated yet.

\begin{figure}[tb!]
\begin{overpic}[width=0.48\textwidth]{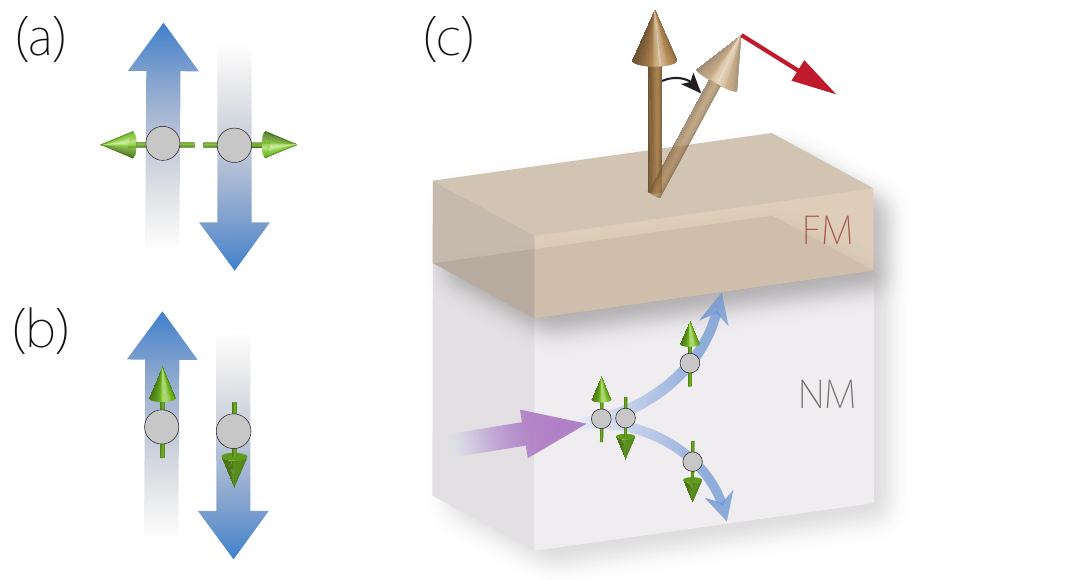}
    \put(55,44){\small\bfseries\color{purple}{$M$}}
    \put(46,16){\small\bfseries\color{violet}{$j$}}
    \put(75,42){\small\bfseries\color{purple}{$M\times(M\times\sigma)$}}
\end{overpic}
    \caption{(a) Spin current induced by conventional spin Hall effect has its spin polarization normal to the flow direction. (b) CPSC has spin polarization collinear with flow direction. (c) A typical device setup. By spin Hall effect, an injected charge current $j$ drives a spin current in the vertical direction, which produces a torque to switch the perpendicular magnetization of the ferromagnetic layer above. To enable field-free switching, it is desired to have CPSC.
%
    \label{fig1}}
\end{figure}

An important motivation for studying spin Hall effect is its potential in manipulating magnetization for information storage and processing \cite{manchon2019,Ryu2020SOT}. A common bilayer device setup is illustrated in Fig.~\ref{fig1}(c), where under a horizontal driving current in the nonmagnetic layer, a vertical spin current resulting from spin Hall effect will flow towards the ferromagnetic layer above and impose a torque to rotate its magnetization. For applications, the magnetic recording layer should have perpendicular magnetic anisotropy, since it has improved storage density, stability, and endurance \cite{manchon2019,Ryu2020SOT,miron2011,Liu2012}. However, switching of perpendicular magnet via spin Hall torques presents a challenge. It was shown that to achieve field-free deterministic switching for perpendicular magnets, the key is to have a collinearly-polarized spin current (CPSC) \cite{kurebayashi2017view,Yang2023briefing}, i.e., having its spin polarization direction collinear with its flow direction (see Fig.~\ref{fig1}(b)), but such a response is prohibited by symmetry in conventional spin source materials, such as heavy metals (see Fig.~\ref{fig1}(a)). In the past few years, a prevailing way to overcome this difficulty is to reduce the symmetry, e.g., by selecting materials with low-symmetry lattice or non-collinear magnetism \cite{Ohno2016,lau2016spin,Ralph2017WTe2,cai2017electric,Baek2018,Nan2020Mn3GaN,kao2022WTe2,Qiu2022Mn3Sn,Wang2023MnPd3,Yang2023TaIrTe4,Yu2023TaIrTe4,Sarkar2024WTe2,Dash2024TaIrTe4,Yang2024,Loh2024}. Despite intensive research, till now, reported materials with
good efficiency in generating CPSC are still limited, which stimulates the exploration of alternative pathways.

In this work, we develop a microscopic theory for the intrinsic NSHE, and reveal that it renders a new way to generate CPSC in a large class of nonmagnetic materials, in which the corresponding linear response is forbidden. We find that CPSC can be induced by NSHE in 18 out of the 21 non-centrosymmetric crystal classes. The intrinsic nonlinear spin Hall conductivity manifests band geometric properties such as Berry-connection polarizability (BCP) and its spin counterpart. Applying our theory to a Weyl model shows that the response can be enhanced by topological nodal features. Combining the theory with first-principles calculations, we conduct quantitative evaluation of intrinsic NSHE in topological metals $\mathrm{PbTaSe_{2}}$ and $\mathrm{PdGa}$, unveiling significant CPSC responses with corresponding spin Hall angle reaching up to $\sim 0.11$ at room temperature.
Our work not only establishes a fundamental intrinsic nonlinear transport phenomenon in spintronics, but also reveals an unexplored route to achieve field-free manipulation of perpendicular magnetization.

{\emph{\textcolor{blue}{CPSC from NSHE.}}}
The nonlinear response of a spin current $j_a^p$ (flowing in direction $a$ with spin polarization $p$) to a driving $E$ field can be expressed as
\begin{equation}\label{chi}
  j_a^p=\sum_{b,c}\chi_{abc}^p E_b E_c,
\end{equation}
where $\chi_{abc}^p$ is the nonlinear spin conductivity tensor. We consider the response in a 3D nonmagnetic bulk material. Then,
$\chi_{abc}^p$ should be a time-reversal-even ($\mathcal{T}$-even) rank-4 pseudo-tensor. One can easily see that like nonlinear charge current, this response also requires broken inversion symmetry. Constraints from other crystalline symmetries can be obtained from the tensor transformation rule:
%
%
%
\begin{equation}
\chi^p_{abc}=\sum_{a'b'c'p'}\text{det}(R)R_{aa^{\prime}%
}R_{bb^{\prime}}R_{cc^{\prime}}R_{pp^{\prime}}\chi^{p^{\prime}}_{a^{\prime}b^{\prime}c^{\prime}}, \label{eq:constraints}%
\end{equation}
with $R$ being a point group operation. From the analysis, we find that this nonlinear response is allowed in all 21 non-centrosymmetric point groups~\cite{supp}.

\begin{table}[ptb]
\setlength\tabcolsep{4.7pt}%
\caption{{}Symmetry-allowed components $\chi_{a(bc)}^{a}$ (corresponding to CPSC) for $\mathcal{T}$-even NSHE in the 9 non-centrosymmetric point groups (PGs) that forbid any CPSC in linear spin Hall response.
For $D_{3h}$, the two-fold axis is
taken along $y$.
}%
\label{symmetry}%
\begin{tabular}{c|c}
\hline \hline
PG & $\chi_{abc}^{a}$\tabularnewline
\hline \hline
$D_{2}$ & $\begin{array}{c}
\chi_{xyy}^{x},\chi_{xzz}^{x},\chi_{yxx}^{y},\chi_{yzz}^{y},\chi_{zxx}^{z},\chi_{zyy}^{z}\end{array}$\tabularnewline
\hline
$C_{2v}$ & $\begin{array}{c}
\chi_{z(xy)}^{z}\end{array}$\tabularnewline
\hline
$D_{4}, D_{6}$ & $\begin{array}{c}
\chi_{xyy}^{x}=\chi_{yxx}^{y},\chi_{xzz}^{x}=\chi_{yzz}^{y},\chi_{zxx}^{z}=\chi_{zyy}^{z}\end{array}$\tabularnewline
\hline
$D_{2d}$ & $\begin{array}{c}
\chi_{xyy}^{x}=-\chi_{yxx}^{y},\chi_{xzz}^{x}=-\chi_{yzz}^{y},\chi_{zxx}^{z}=-\chi_{zyy}^{z}\end{array}$\tabularnewline
\hline
$D_{3h}$ & $\begin{array}{c}
\chi_{y(xz)}^{y}\end{array}$\tabularnewline
\hline
$T$ & $\begin{array}{c}
\chi_{xyy}^{x}=\chi_{yzz}^{y}=\chi_{zxx}^{z},\chi_{xzz}^{x}=\chi_{yxx}^{y}=\chi_{zyy}^{z}\end{array}$\tabularnewline
\hline
$O$& $\begin{array}{c}
\chi_{xyy}^{x}=\chi_{yzz}^{y}=\chi_{zxx}^{z}=\chi_{xzz}^{x}=\chi_{yxx}^{y}=\chi_{zyy}^{z}\end{array}$\tabularnewline
\hline
$T_{d}$ & $\begin{array}{c}
\chi_{xyy}^{x}=\chi_{yzz}^{y}=\chi_{zxx}^{z}=-\chi_{xzz}^{x}=-\chi_{yxx}^{y}=-\chi_{zyy}^{z}\end{array}$\tabularnewline
\hline \hline
\end{tabular}
\end{table}

As mentioned, we are most interested in possible CPSC produced by NSHE. This simply requires $a=p\neq b, c$ in $\chi^p_{abc}$. From Eq.~(\ref{chi}), we may symmetrize the indices $b$ and $c$, and focus on $\chi_{a(bc)}^{a}\equiv(\chi_{abc}^{a}+\chi_{acb}^{a})/2$. From our symmetry analysis, we find that CPSC is widely supported by NSHE: 18 out of the 21 non-centrosymmetric crystal classes can accommodate CPSC response (The exceptions are $C_{3v}$, $C_{4v}$, and $C_{6v}$). The detailed information is presented in \cite{supp}.
Moreover, 9 of these 18 classes, including $C_{2v}$, $D_2$, $D_4$, $D_6$, $D_{2d}$, $D_{3h}$, $T$, $O$ and $T_d$, forbid any CPSC in linear response. In other words, NSHE will produce the leading order contribution to CPSC in these systems. The detailed structures of $\chi_{a(bc)}^{a}$ for these 9 classes are listed in Table~\ref{symmetry}.
This discussion indicates that NSHE greatly broadens the scope of spin current source materials for CPSC.

{\emph{\textcolor{blue}{Theory of intrinsic NSHE.}}}
Next, we develop the microscopic theory of intrinsic NSHE, which is completely determined by the band structure of a material. We adopt the approach of extended semiclassical theory \cite{gao2014,gao2015,gao2019,xiao2021adiabatic}, which has been widely used in studying various nonlinear effects. In this approach, the induced spin current is expressed as
\begin{equation}\label{sc}
  j_a^p=\int [d\bm k] f_{n\bm k} \langle W_{n\bm k}|\hat{j}^p_a |W_{n\bm k}\rangle,
\end{equation}
where $[d\bm k]$ is a shorthand notation for $\sum_n d\bm k/(2\pi)^3$ with $n$ as the band index, $\hat{j}^p_a=\frac{1}{2}\{\hat{v}_a,\hat{s}^p\}$ is the spin current operator,
the last factor is the spin current carried by the wave packet $|W_{n\bm k}\rangle$ which in the absence of $E$ field is centered at Bloch state $|u_{n\bm k}\rangle$, and $f_{n\bm k}$ is the distribution function.
For an intrinsic response, we take Fermi-Dirac distribution for $f_{n\bm k}$, and
the last factor in (\ref{sc}) can be obtained from a variation of semiclassical action, as developed in Refs. \cite{dong2020,xiao2021adiabatic} and detailed in
\cite{supp}.

Via cumbersome but straightforward calculations, we find that the intrinsic nonlinear spin conductivity is given by (we set $e=\hbar=1$)
%
%
%
\begin{equation}\begin{split}\label{key}
  \chi_{a(bc)}^{p}=&\frac{1}{2}\int [d\bm k] \Big[\Lambda_{abc,n}^{p}f_0
  \\
  &-\Big(\langle j_a^p\rangle_n G_{bc,n}-\langle v_b\rangle_n \mathfrak{G}_{ac,n}^p-\langle v_c \rangle_n \mathfrak{G}_{ab,n}^p\Big)f_0'\Big].
  \end{split}
\end{equation}
Here, we suppressed the $\bm k$ labels in the integrand, $f_0\equiv f_0(\varepsilon_n)$ is the Fermi distribution with
$\varepsilon_n$ the unperturbed band energy; $\langle v_a\rangle_n=\langle u_{n}|\hat{v}_a|u_{n}\rangle$ is the intraband velocity matrix element, similar for $\langle j_a^p\rangle_n$.
%
Besides,
\begin{equation}\label{k-BCP}
  G_{bc,n}=2 \text{Re}\sum_{\ell \neq n}\frac{\langle v_b\rangle_{n\ell}\langle v_c\rangle_{\ell n}}{(\varepsilon_{n}-\varepsilon_{\ell})^3}
\end{equation}
is known as the BCP \cite{gao2014,liu2021,liu2022third}, which already plays important roles in several nonlinear effects \cite{Gao2023QM,lai2021}, and
\begin{equation}\label{m-BCP}
  \mathfrak{G}_{ac,n}^p=2 \text{Re}\sum_{\ell\neq n}\frac{\langle j_{a}^{p}\rangle_{n\ell}\langle v_c\rangle_{\ell n}}{(\varepsilon_{n}-\varepsilon_{\ell})^3}
\end{equation}
may be dubbed the spin BCP, in analogy to the concept of spin Berry curvature which is connected to Berry curvature by replacing a velocity matrix element by a spin current matrix element~\cite{Yao2005}.
In the same sense,
\begin{widetext}
\begin{align}
\Lambda_{abc,n}^{p} =-2\mathrm{{\operatorname{Re}}}\sum_{m\neq n}&\bigg[\frac
{3\langle v_{b}\rangle_{nm}\langle v_{c}\rangle_{mn}\left(\langle j_{a}^{p}\rangle_n-\langle j_{a}^{p}\rangle_m \right)
}{(\varepsilon_{n}-\varepsilon_{m})^{4}}
-\frac{\langle \partial_{b} {j}_{a}%
^{p}\rangle_{nm}\langle v_{c}\rangle_{mn}+\langle \partial_{c}{j}_{a}%
^{p}\rangle_{nm}\langle v_{b}\rangle_{mn}}{(
\varepsilon_{n}-\varepsilon_{m})  ^{3}} \nonumber\\
& -\sum_{\ell\neq n} \frac{\left(
\langle v_{b}\rangle_{\ell m}\langle v_{c}\rangle_{mn}+\langle v_{c}\rangle_{\ell m}\langle v_{b}\rangle_{mn}\right) \langle j_{a}^{p}\rangle_{n\ell}}{(  \varepsilon_{n}-\varepsilon_{\ell})  (\varepsilon
_{n}-\varepsilon_{m})^{3}}
-\sum_{\ell\neq m} \frac{\left(
\langle v_{b}\rangle_{\ell n}\langle v_{c}\rangle_{nm}+\langle v_{c}\rangle_{\ell n}\langle v_{b}\rangle_{nm}\right) \langle j%
_{a}^{p}\rangle_{m\ell}}{(  \varepsilon_{m}-\varepsilon_{\ell})  (\varepsilon
_{n}-\varepsilon_{m})^{3}}\bigg]
\end{align}
\end{widetext}
can be regarded as the spin counterpart of BCP dipole $\partial_a G_{bc}$, where $\partial_{a                         }\equiv \partial/\partial{{k}_{a}}$.

We have a few remarks before proceeding. First, the formula of $\chi$ obtained above is for general second-order spin current response, which also includes the longitudinal response. Intrinsic NSHE corresponds to the components with $a\neq b,c$. Second, if all the involved spin current matrix elements are replaced by velocity elements, Eq.~(\ref{key}) will recover the intrinsic nonlinear charge Hall conductivity obtained in previous works \cite{gao2014,wang2021,liu2021}. Meanwhile, when spin $s^p$ is a conserved quantity, one finds that $\chi_{abc}^{p}$ becomes the difference between the intrinsic nonlinear Hall currents, i.e., the antisymmetrized BCP dipoles \cite{liu2021}, between spin-up and spin-down channels. Third, one can verify that the integrand of Eq.~(\ref{key}) is even under $\mathcal{T}$ operation, complying with the $\mathcal{T}$-even nature of intrinsic NSHE. Finally, like Berry curvature, BCP reflects interband coherence (Refs. \cite{wang2021,Gao2023QM,Feng2024QMT} also shows its connection to quantum metric). Hence, we expect the intrinsic NSHE could be enhanced by band (anti-)crossings in the band structure. This is demonstrated in the following.

\begin{figure}[tb!]
\begin{overpic}[width=8.6cm]{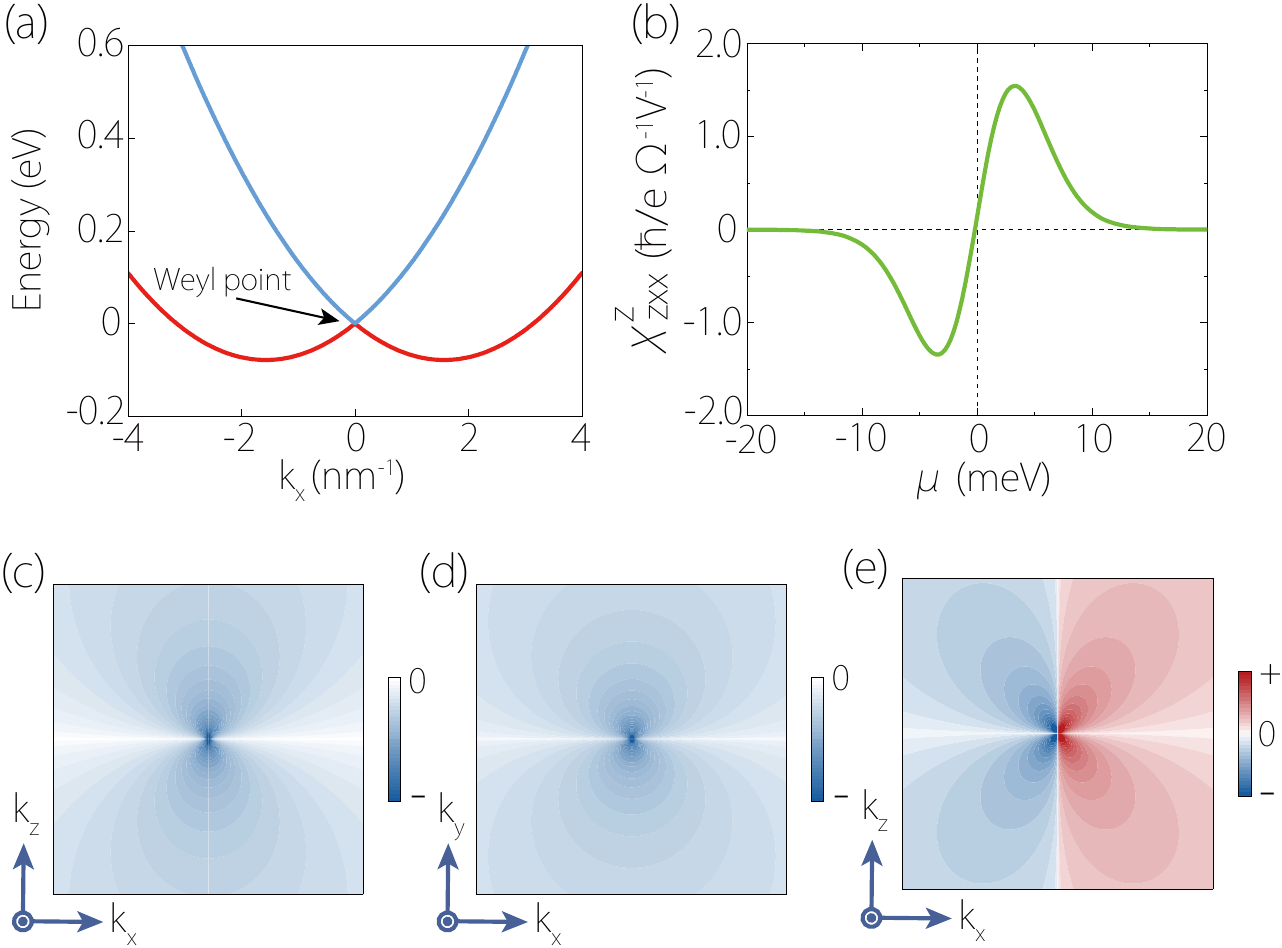}
    \put(13,30){\small\bfseries\color{black}{$\Lambda_{zxx}^{z}$}}
    \put(46,29.9){\small\bfseries\color{black}{$G_{xx}$}}
    \put(79,30.3){\small\bfseries\color{black}{$\mathfrak{G}_{zx}^z$}}
\end{overpic}
    \caption{ (a) Band structure of the Weyl model in Eq.~(\ref{model}). (b) Calculated intrinsic NSHE conductivity $\chi_{zxx}^z$ versus chemical potential. (c-e) show the $k$-space distribution of (c) $\Lambda_{zxx}^{z}$, (d) $G_{xx}$, and (e) $\mathfrak{G}_{zx}^z$ for the lower band. In the calculation, we take $m=1.2 m_e$ ($m_e$ is the bare electron mass), $\nu=1.0\ \text{eV}\cdot \text{\AA}$, and $T = 20$ K.
    \label{fig4}}
\end{figure}

\textcolor{blue}{\textit{Weyl model.}}
To illustrate features of intrinsic NSHE and its involved band geometric quantities, we apply our theory to a simple Weyl model, given by
\begin{equation}
H_{0}(\boldsymbol{k})=\frac{k^{2}}{2m}+\nu \boldsymbol{k} \cdot \boldsymbol{\sigma},
\label{model}
\end{equation}
where $\boldsymbol{\sigma}$ is the vector of Pauli matrices for spin, $m$ and $\nu$ are real model parameters.
This model has two bands with isotropic dispersion, and their crossing forms a Weyl point  at $k=0$ at zero energy (see Fig.~\ref{fig4}(a)).
As an effective model, it
can be realized at $\Gamma$ point for $T$ or $O$ group. From Table~\ref{symmetry}, these groups permit multiple NSHE components. Here, we focus on the component  $\chi_{zxx}^z$, which gives CPSC.

%

The $\chi_{zxx}^z$ obtained from Eq.~(\ref{key}) versus chemical potential $\mu$ is plotted in Fig.~\ref{fig4}(b). The curve has a heartbeat-like shape. One can see that the value is peaked in the energy range around Weyl point and changes sign at the energy of Weyl point.
Furthermore, we derive analytic expressions for the relevant band quantities in $\chi_{zxx}^z$ within this model:
\begin{align}
  G_{xx,\pm}&=\pm \frac{1}{4 \nu}\frac{k^2k_y^2+k_x^2k_z^2 }{k^5 (k_x^2+k_y^2)},
  \\
  \mathfrak{G}_{zx,\pm}^z&=\mp \frac{1}{8m\nu^2}\frac{k_xk_z^2 }{k^5},\\
  \Lambda_{zxx,\pm}^z&=\pm \frac{3}{8m\nu^2}\frac{(k_x^2 k_z^2+k^2k_y^2)k_z^2 }{k^7 (k_x^2+k_y^2)},
\end{align}
where we label the two bands by $\pm$. They exhibit opposite values between upper and lower bands, accounting for the observed heartbeat-like variation in $\chi_{zxx}^z$. While $G_{xx}$ depends on the chirality $\text{sgn}(\nu)$ of Weyl point, $\mathfrak{G}_{zx}^z$ and $\Lambda_{zxx}^z$ are independent of the chirality. In addition, if we kill the quadratic term by letting $m\rightarrow \infty$, $\mathfrak{G}_{zx}^z$ and $\Lambda_{zxx}^z$ will be suppressed in this model, whereas  $G_{xx}$ is not affected. In Fig.~\ref{fig4}(c-e), we plot the $k$-space distribution of these quantities for the lower band. One finds that they are all concentrated around the Weyl point, which conforms with the expectation that as band geometric quantities they manifest interband coherence. Therefore, to enhance NSHE, we may seek materials with band (anti-)crossings around Fermi level. Topological metals could be suitable candidates for this purpose.

\begin{figure}[t!]
\centering
\includegraphics[width=0.48\textwidth]{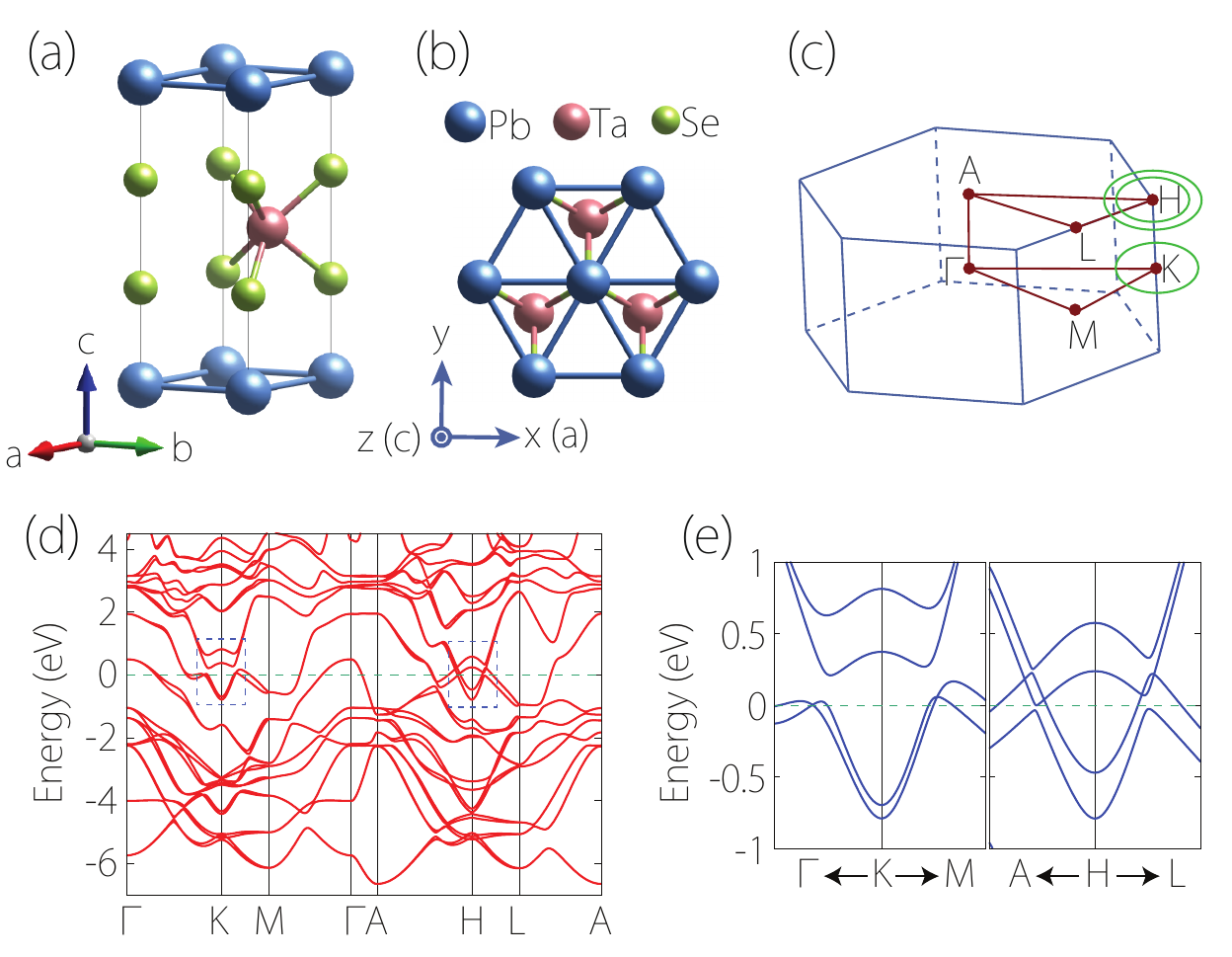} \caption{(a,b) Lattice structure of $\mathrm{PbTaSe_{2}}$. (c) Brillouin zone, with the three nodal rings illustrated by green circles. (d) Calculated band structure. (e) Enlarged view of the blue dashed boxes in (d), showing the band crossings (belonging to the nodal rings) near the Fermi level.
}%
\label{fig2}%
\end{figure}

\textcolor{blue}{\textit{First-principles study of PbTaSe$_2$.}}
As an intrinsic property, the NSHE conductivity in Eq. (\ref{key})  can be readily assessed by first-principles calculations for real materials. We study a concrete example: the nonmagnetic topological nodal-line metal $\mathrm{PbTaSe_{2}}$. It is an existing material, and its topological and charge transport properties have been reported in several previous works~\cite{bian2016,Guan2016,zhang2016,chang2016,pang2016,xu2019}.
The hexagonal lattice structure of $\mathrm{PbTaSe_{2}}$ is shown in Fig.~\ref{fig2}(a-b), which possesses the space group $P\bar{6}m2$ (No.~187) and point group $D_{3h}$. Its symmetry forbids any CPSC response from the linear spin Hall effect.
In contrast, according to Table~\ref{symmetry}, CPSC from NSHE is allowed and is characterized by a single component $\chi_{y(xz)}^{y}$ (the coordinate axis are marked in Fig.~\ref{fig2}(b))~\cite{supp}.

The calculated band structure of $\mathrm{PbTaSe_{2}}$ is plotted in Fig.~\ref{fig2}(d-e), which is consistent with previous results~\cite{bian2016,chang2016,xu2019} (Calculation details are given in \cite{supp}). There are three nodal rings close to Fermi level: One in the $k_{z}=0$ plane around K point and another pair in the $k_{z}=\pi$ plane around H point, as illustrated in Fig.~\ref{fig2}(c) and also shown in Fig.~\ref{fig2}(e). These nodal rings are protected by the mirror symmetry $m_z$. The strong spin-orbit coupling and the topological band structures are promising to trigger large NSHE in $\mathrm{PbTaSe_{2}}$.

\begin{figure}[t]
\centering
\includegraphics[width=0.48\textwidth]{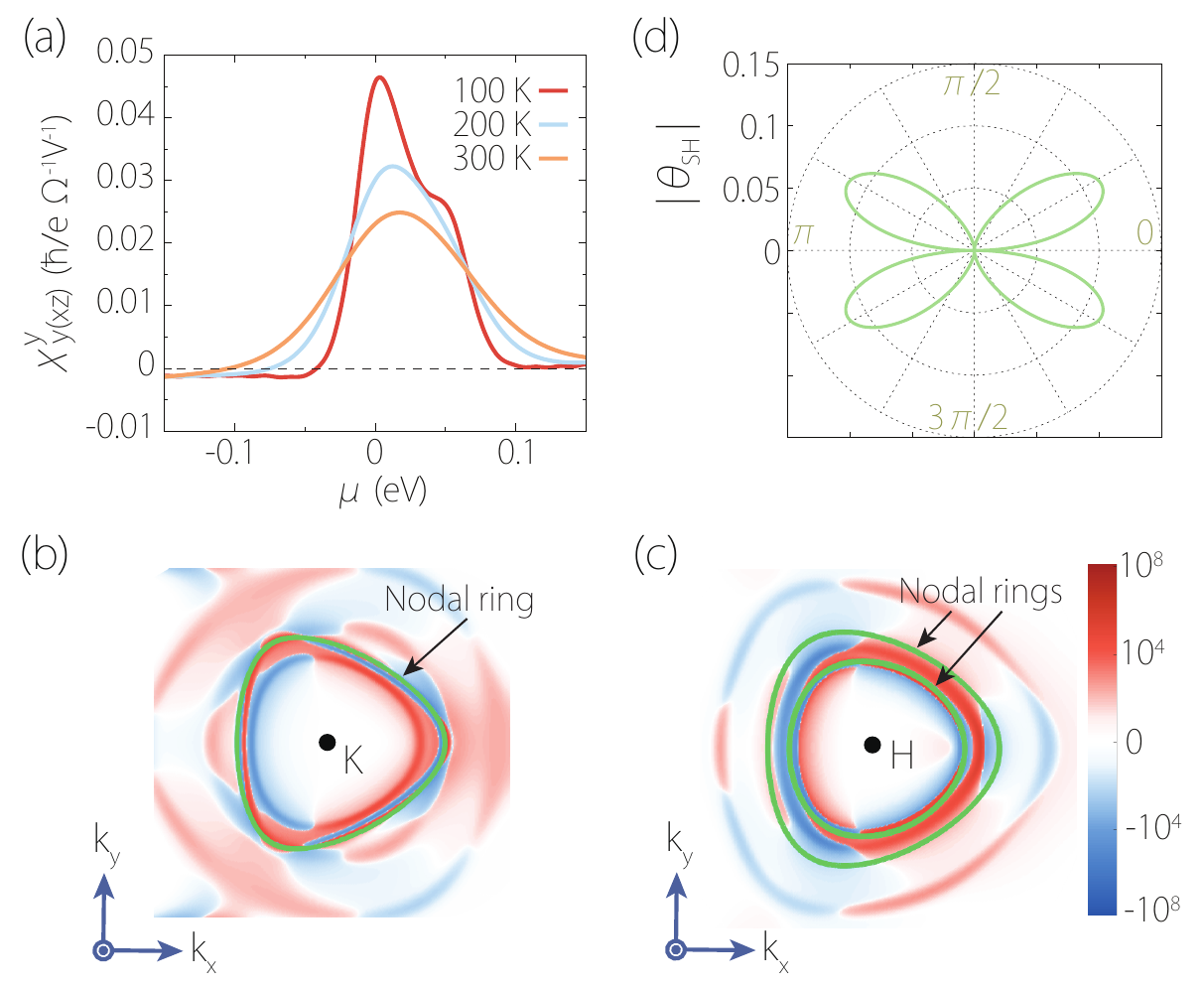}
\caption{(a) Calculated NSHE conductivity $\chi_{y(xz)}^y$ for $\mathrm{PbTaSe_{2}}$ versus chemical potential at different temperatures.
(b,c) $k$-resolved contributions to $\chi_{y(xz)}^y$ on the intrinsic Fermi level (b) in $k_{z}=0$ plane around K point and (c) in $k_{z}=\pi$ plane around H point. The green lines indicate the nodal rings. The unit of color-map is $\mathrm{\AA^{3}/eV}$.
(d) Spin Hall angle versus $\phi$, evaluated at room temperature with  $j$ = $\mathrm{10^6 A/cm^2}$.  }
\label{fig3}%
\end{figure}

Figure~\ref{fig3}(a) shows the calculated $\chi_{y(xz)}^y$ as a function of chemical potential at different temperatures. The response is indeed pronounced when $\mu$ approaches the intrinsic Fermi level, where the nodal rings are located. Particularly, $\chi_{y(xz)}^y=0.046$ $(0.023)$ $\hbar/e$ $\Omega^{-1}$V$^{-1}$ at 100 (300) K for $\mu=0$.
To pinpoint the origin of this large nonlinear response, we plot the $k$-resolved contribution to $\chi_{y(xz)}^y$, i.e., the integrand of Eq.~(\ref{key}), in Fig.~\ref{fig3}(b-c). One observes that the large contributions are concentrated near the nodal rings (green lines). More precisely, around H point [Fig.~\ref{fig3}(c)], the inner nodal-ring region is more dominant as it is closer to Fermi level than the outer ring.
These results demonstrate that the NSHE response is greatly enhanced by topological band features in $\mathrm{PbTaSe_{2}}$.

The charge-to-spin current conversion efficiency is usually characterized by the spin Hall angle.
For NSHE, the effective spin Hall angle can be defined as $\theta_{\mathrm{SH}}=(2e/\hbar)\chi\rho^{2}j$, where $j$ is the injected charge current density, and $\rho$ is the longitudinal resistivity. For spin-orbit torque applications as in Fig.~\ref{fig1}(c), $\theta_{\mathrm{SH}}$ is also a measure of the spin torque efficiency \cite{manchon2019}. When the charge current is injected within the $zx$ plane making an angle $\phi$ with the $z$ axis, one has  $\chi=\chi^y_{y(xz)}\sin 2\phi$, and $\rho=\rho_{xx}\sin^2 \phi + \rho_{zz}\cos^2\phi$. Here we take $\rho_{xx}\sim \mathrm{0.57\ m\Omega\  cm}$ from Ref.~\cite{zhang2016} and estimate $\rho_{zz}\sim 3.62 \rho_{xx}$ from the Drude weight calculation. Taking a relatively small charge current $j=\mathrm{10^6\ A/cm^2}$ (in usual spin Hall torque measurements, $j$ locates within $10^6 \sim 10^8$ $\mathrm{A/cm^2}$ \cite{manchon2019}), we plot in Fig.~\ref{fig3}(d) the angular dependence of $|\theta_{\mathrm{SH}}|$ versus $\phi$ at $T=300$ K. One observes that $\theta_{\mathrm{SH}}$ can reach up to $\sim 0.11$ at $\phi\approx 27^\circ$, which is already comparable to the largest out-of-plane linear spin Hall torque efficiency (0.03 $\sim$ 0.11) reported in low-symmetry materials~\cite{Yang2023TaIrTe4,Yu2023TaIrTe4,Dash2024TaIrTe4}.

\begin{figure}[tb!]
\begin{overpic}[width=0.45\textwidth]{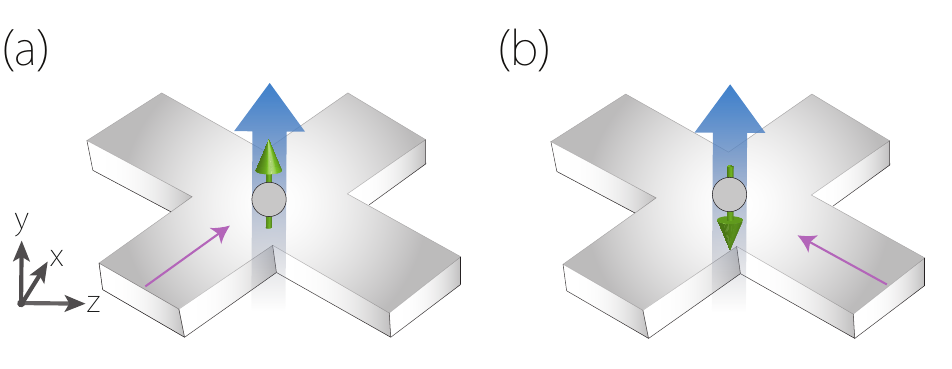}
    \put(22,32){\small\bfseries\color{blue}{$j_{y}^{y}=+$}}
    \put(70,32){\small\bfseries\color{blue}{$j_{y}^{y}=-$}}
    \put(13,5.5){\small\bfseries\color{violet}{$j$}}
    \put(8,1.5){\small\bfseries\color{violet}{$\phi=45^{\circ}$}}
    \put(92.5,5.5){\small\bfseries\color{violet}{$j$}}
    \put(86.5,1.5){\small\bfseries\color{violet}{$\phi=135^{\circ}$}}
\end{overpic}
    \caption{Schematic figure showing that generated CPSC (and the torque) can be readily switched by controlling
    the driving current direction. Here shows two representative cases with (a) $\phi = 45^\circ$ and (b) $\phi = 135^\circ$.
    \label{fig5}}
\end{figure}
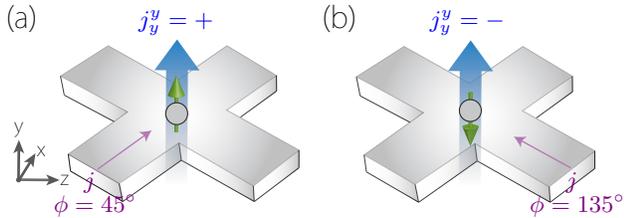

In addition, the sign of the nonlinear spin Hall torque can be effectively switched by controlling the driving current direction in $\mathrm{PbTaSe_{2}}$. The induced CPSC $j^y_y=\chi^y_{y(xz)} E^2\sin 2\phi$ has a period of $\pi$ with respect to current injection angle $\phi$. It
is positive for $\phi\in(0,\pi/2)\cup(\pi, 3\pi/2)$, negative for $\phi\in(\pi/2,\pi)\cup(3\pi/2,2\pi)$, and vanishes at the boundaries between the four quadrants. Hence, one can readily switch the generated torque by varying $\phi$,
as illustrated in Fig.~\ref{fig5}. These results suggest $\mathrm{PbTaSe_{2}}$ holds great potential as a nonlinear spin source material for field-free control of perpendicular magnets.

\textcolor{blue}{\textit{Discussion.}}
We have established the theory of intrinsic NSHE and revealed NSHE as a new route to generate CPSC for switching of perpendicular magnets. The strong NSHE predicted in $\mathrm{PbTaSe_{2}}$ can be readily probed in experiment, e.g., by magneto-optical Kerr measurement \cite{Fan2014,Wang2015MOKE}, or spin-torque ferromagnetic resonance measurement \cite{vyborny2011,Yang2023TaIrTe4}, which can resolve torque components
along different directions. The distinct $\phi$-dependence discussed above (Fig.~\ref{fig5}) also makes it easy to distinguish the effect from any linear torque response (which must flip sign between $\phi$ and $\phi+\pi$).

Our work greatly broadens the scope of spin source materials for CPSC. Materials with forbidden linear CPSC response, which hence is left out of consideration previously, may possess strong nonlinear response from NSHE. $\mathrm{PbTaSe_{2}}$ is such an example. In the Supplemental Material~\cite{supp}, we present another example PdGa, which has CPSC response of a similar magnitude as $\mathrm{PbTaSe_{2}}$.


Finally, our study can be naturally extended to magnetic materials, in which both $\mathcal{T}$-even and $\mathcal{T}$-odd NSHE may be allowed (From symmetry, $\mathcal{T}$-odd NSHE must be of extrinsic character). There, the interplay between magnetic ordering and NSHE would be an interesting topic to explore and may lead to new functionalities.


\bibliographystyle{apsrev4-2}
\bibliography{ref}

\end{document}